\documentclass[preprint,12pt]{elsarticle}

\usepackage{amssymb}

\usepackage{float}

\journal{Chaos, Solitons $\&$ Fractals}

\begin{document}

\begin{frontmatter}



\title{Selectivity filter conductance, rectification and fluctuations of subdomains - how can this all relate to the value of Hurst exponent in the dwell-times of ion channels states?}

\author[inst1]{Przemysław Borys}\corref{cor1}
\ead{przemyslaw.borys@polsl.pl}
\affiliation[inst1]{organization={Department of Physical Chemistry and Technology of Polymers, Silesian University of Technology},
            addressline={Strzody 9}, 
            city={Gliwice},
            postcode={44-100}, 
            country={Poland}}
            
\author[inst2]{Paulina Trybek}
\affiliation[inst2]{organization={Institute of Physics, University of Silesia},
            addressline={Pułku Piechoty 1}, 
            city={Chorzów},
            postcode={41-500}, 
            country={Poland}}

\author[inst3]{Beata Dworakowska}

\affiliation[inst3]{organization={Institute of Biology, Department of Physics and Biophysics, Warsaw University of Life Sciences},
            addressline={Nowoursynowska 159}, 
            city={Warsaw},
            postcode={02-787}, 
            country={Poland}}
            
\author[inst3]{Anna Sekrecka-Belniak}
\author[inst4]{Ewa Nurowska}
\affiliation[inst4]{organization={Laboratory of Physiology and Pathophysiology, Centre for Preclinical Research and Technology, Faculty of Pharmacy, Medical University of Warsaw},
            addressline={Banacha 1B}, 
            city={Warsaw},
            postcode={02-097}, 
            country={Poland}}

\author[inst3]{Piotr Bednarczyk}
\author[inst1]{Agata Wawrzkiewicz-Jałowiecka}\corref{cor2}
\ead{agata.wawrzkiewicz-jalowiecka@polsl.pl}

\begin{abstract}
The Hurst effect in the signals describing ion channels' activity has been known for many years. This effect is present in the experimental recordings of single-channel currents, but not only. The sequences of dwell times of functionally different channel states also exhibit long-range correlations.
We have found that the memory effect within the dwell-time series is related to the coupling between the channel's activation gate (AG) and selectivity filter (SF), which controls the ion conduction.
In this work, we analyzed both the experimental data describing the activity of potassium channels of different types (e.g., BK, mitoBK, mitoTASK-3,  mitoKv1.3, TREK-2-like channels) and the series generated according to our previously proposed Hurst effect model.  
The obtained results suggest that the strength of the allosteric cooperation between the AG and SF determines not only the conductance of the channel - which governs how often ions in SF move or remain blocked - but also modulates the correlations present in the dwell times when sampled with a suitably high sampling rate. Moreover, we found that rectification can interfere with this process, contributing to additional changes in correlations within the channel's sojourns in subsequent states.
Similarly, the correlations may be affected by processes proceeding at longer time scales, like interactions with the channel's auxiliary domains or lipid surroundings.
\end{abstract}




\end{frontmatter}


\section{Introduction}

The activity of ion channels can be considered as a complex, noisy process that deeply affects the biology of virtually all known cells~\cite{hille2001ion}. Throughout decades, plethora of investigations has been carried out in aim to improve our understanding of different aspects of ion channel functioning.
One of the compelling features of the signals describing ion channel gating is the presence of the long-range correlations (memory)~\cite{silva2021memory}.
This effect was described for the first time in~\cite{nogueira1995hurst}. In that work, the authors evaluated the Hurst R/S (rescaled range) exponent~\cite{hurst1956problem} to study the correlations within the dwell-time series of open and closed states of the $Ca^{2+}$-activated $K^+$ channels of Leydig cells. Their discovery of the existence of long-range correlations in analyzed data, started the still open debate whether channel gating is a purely stochastic process or it also possesses some deterministic components. This is because the simple Markov process, which has been used to describe gating kinetics for a long time, is not able to reproduce the memory effect~\cite{nogueira1995hurst}.

Up to date, many studies investigate the memory in ion channel activity in the series of single-channel currents and/or the corresponding dwell-time series 
\cite{nogueira1995hurst, kochetkov1999non, varanda2000hurst, siwy2001application, siwy2002correlation, lan2003rescaled,  de2006long, lan2008detrended, peng2012existence, wawrzkiewicz2012simple, wawrzkiewicz2017temperature, m2017dynamic, wawrzkiewicz2018mechanosensitivity, wawrzkiewicz2020differences, miskiewicz2020long, wawrzkiewicz2021dynamical}, as summarized in Table~\ref{Tab:Hurst_literature}. In these investigations, the Hurst exponent (H) 
being the measure of predictability of the series is evaluated, according to either the original R/S algorithm~\cite{hurst1956problem} or the Detrended Fluctuation Analysis (DFA)~\cite{peng1994mosaic}.
In literature, there are proposed dynamical models of channel gating that can reproduce the Hurst effect~\cite{bandeira2008chaotic, wawrzkiewicz2012simple, bahramian2019introducing, borys2020long}.

\begin{table}[H]
\caption{Summary of the Hurst memory analysis in ion channel studies in the literature. For each reference, the applied method for Hurst exponent ($H$) evaluation is indicated (i.e., the R/S algorithm or the DFA). Please note, that these cited results may depend much on the length of the samples considered.\\} 
\label{Tab:Hurst_literature}
\centering
\begin{tabular}{p{0.17\textwidth}p{0.23\textwidth}p{0.22\textwidth}p{0.1\textwidth}p{0.33\textwidth}}
\hline
\textbf{Reference} & \textbf{Channel/model type} & \textbf{Analyzed data type} & \textbf{H evaluation method} & \textbf{Memory effect, H values}\\
\hline

   Nogueira et al. (1995) \cite{nogueira1995hurst} & $Ca^{2+}$-activated $K^+$ channels in Leydig cells & dwell-time series & R/S & trend-reinforcing; $H_{RS}$ = $\sim$0.75\\
   & simulated series according to Markovian model & dwell-time series & R/S & uncorrelated\\
\hline
   Kochetkov et al. (1999) \cite{kochetkov1999non} & BK channels in kidney cells & dwell-time series & R/S & trend-reinforcing; $H_{RS}$=0.60$\pm$0.04 in the short temporal regime, and $H_{RS}$ = 0.88$\pm$0.21 in the long temporal regime
   \\
   & simulated series according to Markovian model & dwell-time series & R/S & uncorrelated\\
\hline
   Varanda et al. (2000) \cite{varanda2000hurst} & $Ca^{2+}$-activated $K^+$ channels in Leydig cells & dwell-time series & R/S & trend-reinforcing; 0.61$\leq$$H_{RS}$$\leq$0.64, non-monotonic dependence on membrane potential\\
   & simulated series according to Markovian model & dwell-time series & R/S & uncorrelated\\ 

\hline
   Siwy et al. (2001) \cite{siwy2001application} & BK channel in extensor tibiae fibers from locust & dwell-time series of open-only, closed-only states & R/S & trend-reinforcing; open states: $H_{RS}$ = 0.67$\pm$0.06, closed states: $H_{RS}$ = 0.62$\pm$0.06\\
   & & single channel currents & R/S & trend-reinforcing; $H_{RS}$ = 0.84$\pm$0.08\\
  
\hline
    
\end{tabular}
\end{table}

\begin{table}[H]
\centering
\begin{tabular}{p{0.17\textwidth}p{0.23\textwidth}p{0.22\textwidth}p{0.1\textwidth}p{0.33\textwidth}}

\hline
    & &  dwell-time series of open-only, closed-only and all states & DFA & trend-reinforcing; open states: $H_{DFA}$ = 0.62$\pm$0.06, closed states: $H_{DFA}$ = 0.57$\pm$0.05 in the short temporal regime and $H_{DFA}$ = 0.86$\pm$0.07 in the long temporal regime, all states: $H_{DFA}$ = 0.58$\pm$0.07 in the short temporal regime and $H_{DFA}$ = 0.85$\pm$0.09 in the long temporal regime\\ 
\hline
   Siwy et al. (2002) \cite{siwy2002correlation} & BK channel in extensor tibiae fibers from locust & cumulative ion current time series corresponding to open-only, closed-only and all states & DFA & trend-reinforcing; open states: $H_{DFA}$ = 0.65$\pm$0.01 in the short temporal regime and $H_{DFA}$ = 0.71$\pm$0.01 in the long temporal regime, closed states: $H_{DFA}$ = 0.69$\pm$0.02 in the short temporal regime and $H_{DFA}$ = 0.98$\pm$0.02 in the long temporal regime, all states: $H_{DFA}$ = 0.83$\pm$0.01 in the short temporal regime and $H_{DFA}$ = 1.04$\pm$0.02=4 in the long temporal regime\\
\hline
   Lan et al. (2003) \cite{lan2003rescaled} & Delayed rectifier $K^+$ channel in rat dorsal root ganglion neurons & dwell-time series & R/S & anti-persistent; 0.34$\leq$$H_{RS}$$\leq$0.40 for different voltages\\
   & Simulated data according to the Markovian model & dwell-time series & R/S & anti-persistent; $H_{RS}$ = 0.25$\pm$0.01\\
\hline
   Campos de Oliveira et al. (2006) \cite{de2006long} & $Ca^{2+}$-activated $K^+$ channels in Leydig cells & dwell-time series & R/S & trend-reinforcing; $H_{RS}$ = 0.64$\pm$0.06\\
   & simulated series according to Markovian model & dwell-time series & R/S & similar to randomized data, $H_{RS}$ = 0.57$\pm$0.03\\
   & simulated series according to fractal model & dwell-time series & R/S & similar to randomized data, $H_{RS}$ = 0.56$\pm$0.02\\
\hline
    
\end{tabular}
\end{table}

\begin{table}[H]
\centering
\begin{tabular}{p{0.17\textwidth}p{0.23\textwidth}p{0.22\textwidth}p{0.1\textwidth}p{0.33\textwidth}}

\hline
   Lan et al. (2008) \cite{lan2008detrended} & $K^+$ channels of 79 [pS] single-channel conductance from rat dorsal root ganglion neurons & single-channel currents & DFA & trend-reinforcing; 0.94$\leq$$H_{DFA}$$\leq$0.97 for different voltages\\
\hline
   Wawrzkiewicz et al. (2012) \cite{wawrzkiewicz2012simple} & BK channels in Human Bronchial Epithelial (HBE) cells & dwell-time series & R/S & trend-reinforcing; 0.63$\leq$$H_{RS}$$\leq$0.82, non-monotonic dependence on membrane potential\\
   & simulated data (Random walk models) & dwell-time series &	R/S & trend-reinforcing; 0.73$\leq$$H_{RS}$$\leq$0.82, non-monotonic dependence on drift force\\
\hline
   Peng et al. (2012) \cite{peng2012existence} & voltage-dependent $K^+$ channel in L$\beta$T2 cells& 0–1 series corresponding to closed and open channel currents & DFA & trend-reinforcing; 0.90$\leq$$H_{DFA}$$\leq$0.98 for different voltages\\
\hline
   Wawrzkiewicz-Jałowiecka et al. (2017) \cite{wawrzkiewicz2017temperature} & BK channels in glioblastoma U-87 MG cells & dwell-time series &	R/S & trend-reinforcing; $\sim$0.66$\leq$$H_{RS}$$\leq$$\sim$0.72, non-monotonic dependence on temperature\\
\hline
   De la Fuente et al. 2017 \cite{m2017dynamic} & $Ca^{2+}$-activated $Cl^-$ channels in Xenopus laevis oocytes & chloride currents & DFA & $H_{DFA}$ = 0.93$\pm$0.05\\
   & & long time intervals of ionic conduction  & R/S & anti-persistent; $H_{RS}$ = 0.19$\pm$0.10\\
   \hline

   Wawrzkiewicz-Jałowiecka et al. (2018) \cite{wawrzkiewicz2018mechanosensitivity} & BK channels in glioblastoma U-87 MG cells & dwell-time series & R/S and DFA & trend-reinforcing; 0.58$\leq$$H_{RS}$$\leq$0.63 and 0.58$\leq$$H_{DFA}$$\leq$0.66 at different strength of mechanical stimulation of the membrane\\
\hline
\end{tabular}
\end{table}

\begin{table}[H]
\centering
\begin{tabular}{p{0.17\textwidth}p{0.23\textwidth}p{0.22\textwidth}p{0.1\textwidth}p{0.33\textwidth}}

\hline
   Wawrzkiewicz-Jałowiecka et al. (2020) \cite{wawrzkiewicz2020differences} & BK channels in glioblastoma U-87 MG cells & dwell-time series for open-only, closed-only and all states & R/S and DFA & trend-reinforcing (in general); open states: $H_{RS}$= 0.59, $H_{DFA}$= 0.72, closed states: $H_{RS}$= 0.53, $H_{DFA}$= 0.83, all states: $H_{RS}$= 0.57, $H_{DFA}$= 0.68\\
   & mitoBK channels in glioblastoma U-87 MG cells & dwell-time series for open-only, closed-only and all states & R/S and DFA & trend-reinforcing (in general); open states: $H_{RS}$= 0.54, $H_{DFA}$= 0.63, closed states: $H_{RS}$= 0.53, $H_{DFA}$= 0.73, all states: $H_{RS}$= 0.52, $H_{DFA}$= 0.62\\
   & BK channels in glioblastoma U-87 MG cells & single-channel currents for open-only, closed-only states & R/S and DFA & trend-reinforcing; open states: $H_{RS}$= 0.67, $H_{DFA}$= 0.73, closed states: $H_{RS}$= 0.73, $H_{DFA}$= 0.81\\
   & mitoBK channels in glioblastoma U-87 MG cells & single-channel currents for open-only, closed-only states & R/S and DFA & trend-reinforcing; open states: $H_{RS}$= 0.68, $H_{DFA}$= 0.86, closed states: $H_{RS}$= 0.72, $H_{DFA}$= 0.94\\
\hline
   Miśkiewicz et al. 2020 \cite{miskiewicz2020long} & nonselective slow activated cationic channels (SV) from Beta vulgaris vacuoles & dwell-time series & R/S and DFA & trend-reinforcing; $H_{RS}$ = 0.85$\pm$0.03 and $H_{DFA}$ = 1.05$\pm$0.07
   \\
\hline
   Wawrzkiewicz-Jałowiecka et al. 2021 \cite{wawrzkiewicz2021dynamical} & mitoBK channels in endothelial cells (EA.hy926) & single channel currents & R/S and DFA & trend-reinforcing; $\sim$0.73$\leq$$H_{RS}$$\leq$$\sim$0.88 and $\sim$0.68$\leq$$H_{DFA}$$\leq$$\sim$0.86, non-monotonic dependence on voltage\\
   & mitoBK channels in dermal fibroblasts (HDFa) & single channel currents & R/S and DFA & trend-reinforcing; $\sim$0.70$\leq$$H_{RS}$$\leq$$\sim$0.76 and $\sim$0.61$\leq$$H_{DFA}$$\leq$$\sim$0.83, non-monotonic dependence on voltage\\
   & mitoBK channels in glioblastoma cells (U-87 MG) & single channel currents & R/S and DFA & trend-reinforcing; $\sim$0.69$\leq$$H_{RS}$$\leq$$\sim$0.86 and $\sim$0.78$\leq$$H_{DFA}$$\leq$$\sim$0.97, non-monotonic dependence on voltage\\ 
\hline
\end{tabular}
\end{table}

In this work we would like to answer the question about the leading mechanistic factors that determine the values of Hurst exponent for the series of dwell-times of channel states. Which factors interact to produce correlations, and which destroy them?
Our initial observations revealed that strong Hurst effect is present in the big conductance $K^+$ channels, which was extensively studied in last years (Table \ref{Tab:Hurst_literature}).
In contrast, the long-range memory effect in $K^+$ channels of lower conductance was barely analyzed hitherto.
For a representative of small-conductance channel, i.e., the delayed rectifier $K^+$ channel~\cite{lan2003rescaled}, the anti-persistent characteristics of gating has been recognized.
It inspired us to consider, what kind of mechanistic component of the channel gating machinery could be responsible for changes in the range from persistent to anti-persistent dynamics when one potassium channel type is compared to another.
Moreover, former results revealed that some differences between the Hurst exponents occur both at different levels of channel's activation (e.g., for the voltage-dependence~\cite{varanda2000hurst, lan2003rescaled, lan2008detrended, wawrzkiewicz2012simple, peng2012existence, wawrzkiewicz2021dynamical}), for various channel variants (e.g., BK and mitoBK~\cite{wawrzkiewicz2020differences}), and when the channels are located in different cell types and coupled with different auxiliary regulating $\beta$ subunits (e.g., \cite{wawrzkiewicz2021dynamical}).

\begin{figure}[H]
\centering
\includegraphics[width=0.7\linewidth]{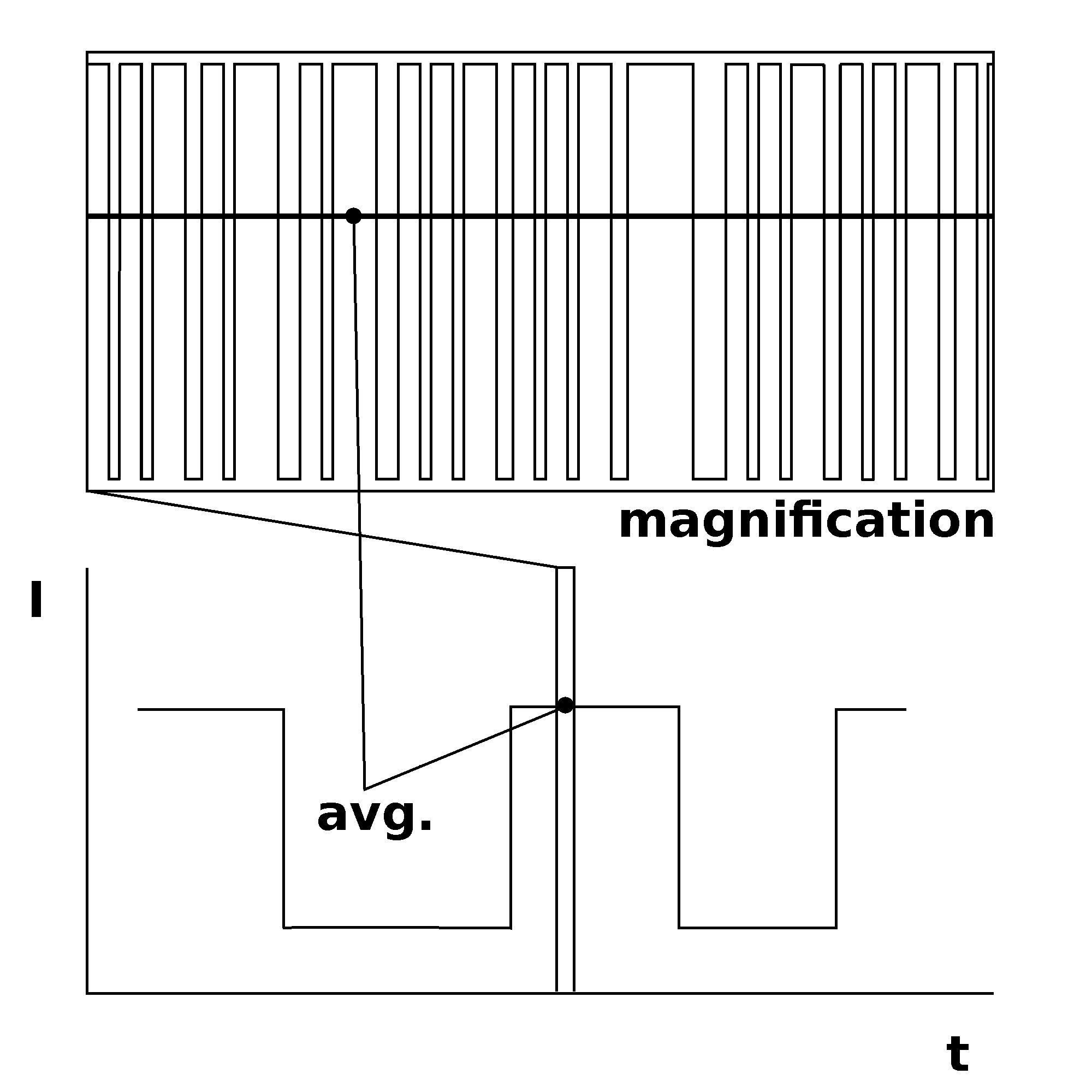}
\caption{Illustration of the impact of selectivity filter fluctuations on the channel's conductance. Majority of the short-living fluctuations of the selectivity filter is too fast for the sampling devices to be caught. Instead, they record an averaged value of the current (thick line in upper panel, denoted ,,avg.''). The measured conductance depends then upon the number and duration of the fast ,,off'' events in relation to the ,,on'' events. In this research we consider a situation, when fluctuations happen to be long enough to be caught by the sampler. If this happens, they can shape the correlation properties of ion channel.}
\label{conductanceillustration}
\end{figure}

Considering the nature of channel transport, its rate is limited by the single file trespass of the ions through the SF (Selectivity Filter), which has a highly conserved structure in all potassium channels. Typically, it takes the form of TVGYG motif~\cite{doyle1998structure}.
The conductance drop in different $K^+$ channel types can only be attributed to fluctuation-driven openings and closings of the SF, which in some species are easier and more frequent than in the others.
These rapid fluctuations become averaged for perturbation dwell-times below the sampling rate, which results just in establishing some net conductance of a channel in the patch-clamp recording (Fig. \ref{conductanceillustration}).
However, especially in channels with frequent fluctuations perturbing the open state (i.e. low conductance channels), occasionally there may occur fluctuations which are long enough to be caught by the sampling device. They will interfere with correlations of the dwell-time signal -- stemming from the AG (Activation Gate) dynamics and the AG-SF coupling -- possibly destroying it.

The picture above is supported by molecular dynamics studies of the selectivity filter gating \cite{kopec2019molecular}, where the filter may be blocked or unblocked depending on nanoscale translations of the T59 residue of the SF. Additionally, the ionic transport may be blocked by water entry into the filter (also facilitated by particular T59 orientations).
The patterns within the sequences of the SF opening and blocking are also modulated by the interactions with the channel pore, various AG domains, and regulatory subunits (RS), which may impose tension on parts of the SF or provide constraints to its fluctuations. The structure and function of these parts of channel protein define the identity of particular channel types \cite{mironenko2021persistent}.
The activation gates and the regulatory subunits can not be considered rigid structures, but rather they fluctuate on their own time scale, much larger than the scale of individual residue fluctuations in the selectivity filter. Therefore, their coupling with the SF leads to additional components in the channel gating pattern and is supposed to affect the existence of correlations within the dwell times of conducting and non-conducting channel states (in fact such interaction of processes with different time scales was proposed to generate correlations, as considered in our model \cite{wawrzkiewicz2012simple}).

From this perspective, in this work, we decided to evaluate and compare the Hurst exponents for the dwell-times series describing the activity of various $K^+$ channel types, which substantially differ by their structures, and, consequently, the single channel conductance, which was the factor that initially brought our high attention.
We also analyzed the more subtle effects on long-range memory, which could be exerted by channel rectification, cooperation of the channel gate with the regulatory subunits, and the presence of distinct channel isoforms in different biological membranes. 
To simplify the interpretation of the results obtained for the experimental series, we have carried out the appropriate simulations, which illustrate the effects of channel rectification and sensitivity of AG-SF dynamics to the SF fluctuations that mirror the conductance properties.


\section{Materials and Methods}
\subsection{Hurst exponent}

\subsubsection{R/S analysis}
The Rescaled Range analysis (R/S) is one of the simplest technique for estimation of Hurst exponent. The Hurst exponent is obtained in the following way \cite{hurst1956problem}:

\begin{itemize}
\item Divide the time series $x_i$ into intervals of length $n$, $x_{ij}$.
\item In each of the sub-series, find the average $\bar x_j$
\item For each sub-series $x_{ij}$ construct $t_{ij}=x_{ij}-\bar x_j$
\item Calculate the standard deviation of $t_{ij}$, and call it $S_j$
\item Construct a series that resembles trajectory performed by steps $t_{ij}$, $y_{0j}=t_{0j}$, $y_{ji}=y_{i-1j}+t_{ij}$ for $i \ge 1$
\item Denote $y_{maxj}-y_{minj}=R_j$
\item This allows to find the ratio of $R_j/S_j$ for a particular sub-series $x_{ij}$. Since $j$ can take a range of values (number of sub-series of length $n$ in the original series) - this quantity can be averaged over $j$ to give final $R/S$ representative for given sub-series length $n$.
\item For various $n$, plot the chart of $R(n)/S(n)$ as function of $n$
\item The scaling is $R/S \sim n^H$ (or $log R/S=H log n + const$), where $H$ is the Hurst coefficient.
\end{itemize}

\subsubsection{DFA analysis}
The idea of Detrended Fluctuation Analysis is based on the assumption that the dynamic of time series is affected by the short-term and long-term features. The technique was first proposed by Peng et al. in 1994 for the identification of correlation in DNA structure at multiple scales \cite{peng1994mosaic}.   
The simple procedure of DFA starts with the calculation of the cumulative signal $y_i$ (\ref{dfa_profile}), where the average value of series $\overline{x}$ is subtracted from the cumulative sum of data $x_i$.

 \begin{equation}
    y_i=\sum_{k=1}^i[x_k-\overline{x}]
    \label{dfa_profile}
\end{equation}

At the next stage, the estimated profile $y_i$ is divided into $N_s$ non--overlapping segments expressed as successive powers of 2. For all segments $v$ ($v = 1,..., N_s)$ the local trend $y_{v,i}^m$ is estimated by the least-square fit with order $m$. The key step of the method involves determining the variance $F^2$ as a function of scale $s$ (the size of the segment v) (\ref{scaling_f}). 

\begin{equation}
    F^2(s,v)=\frac{1}{s} \sum_{i=1}^s(y_{v,i}-y_{v,i}^m)^2
    \label{scaling_f}
\end{equation}

At the final stage the Hurst exponent $H$ is estimated as the slope of the regression line of double-logarithmic dependence, $log(F)$ in the function of $log(s)$ ($log(F) \approx H log(s)$).

\subsection{Empirical dwell-time series}

Dwell-time series analyzed in this work are obtained for the experimental time series of the single-channel currents recorded by the use of the patch-clamp method on the excised patches of biological membranes. 
The following channel types are investigated: mitochondrial TASK-3 channels from human keratinocyte HaCaT cells (mitoTASK-3), the voltage-gated Kv1.3 channels from the inner mitochondrial membrane of gerbil hippocamp cells (mitoKv1.3), TREK-2-like channels from pyramidal neurons of rat prefrontal cortex (TREK-2), and the large-couductance voltage- and $Ca^{2+}$-activated $K^+$ channel (BK) channels from different cell lines as well as their mitochondrial variants (mitoBK).
The detailed description of the experimental procedures used for cell culture, mitochondria and mitoplast preparation (in the case of mitochondrial channels), as well electrophysiological recordings is provided in appropriate original studies, from which the selected recordings have been adapted for this research; i.e., mitoTASK-3 at membrane potential U = -90 mV and 90 mV \cite{toczylowska2014potassium}, mitoKv1.3 at membrane potential U = 40 mV \cite{bednarczyk2010identification}, TREK-2-like channels U = 50 mV \cite{dworakowska2021activity}, BK and mitoBK from glioblastoma U87-MG cells \cite{bednarczyk2013putative, wawrzkiewicz2020multifractal}, BK channels from human bronchial epithelial cells at membrane potential U = 50 mV and three concentrations of $Ca^{2+}$ ions (0, 10 and 100 $\mu$M) \cite{wawrzkiewicz2012simple}, and BK and mitoBK from human endothelial cell line (EA.hy926) at U = 40 mV \cite{bednarczyk2013large}.
The samples of experimental recordings are presented in \ref{Appendix A} in Figs. \ref{figA1}, \ref{figA2}, \ref{figA3}.
For each from the compared groups of channels, we selected at least 3 independent dwell-time series of N = 2100 states. Only in the case of the mitoTASK-3, the analyzed series were shorter (N ranged from 262 to 2100) due to the difficulties with electric stability of a patch during the recording.
In turn, for the BK channels from glioblastoma, the relatively long measurements were possible, and allowed us to obtain the series of 37000 to 212000 sojourns.

\subsection{Simulated series}

We have revisited the random walk model of channel gating presented in \cite{wawrzkiewicz2012simple} to explain the effects of differences in channel surroundings on the conformational diffusion of the gate (like in the case of the plasma membrane- and mitochondrial channels located in different lipid microenvironment).
In short, the model assumes that the channel gating can be illustrated as a discrete random walk process performed by the reaction coordinate (RC) over the one-dimensional conformational space (discrete lattice).
The size of the diffusive space of the RC is limited by the position of two movable boundaries (B1 and B2) symmetrically located around the threshold point (TP) that separates open and closed states (TP = 0; B1=-B2; if RC$<$TP, the closed state is recognized; if RC$>$TP, the open state is recognized; RC=TP not allowed). The fluctuations of B1 and B2 are synchronized in direction (e.g., when the conformational space corresponding to the open state shrinks, the size of the closed conformational space also decreases).
The probability of increasing or reducing the accessible conformational space by the boundary motion was equal to 0.5. The maximal position of the boundary is 2$B_{max}$. (Here, the $B_{max}$ = 14.)
Because the RC corresponds to the channel gate and the boundaries correspond to its surroundings, which significantly differ by mass and propensity to thermal fluctuations, the random motion of B1 and B2 should be realized on a larger time scale ($\tau_B$) than the jumps of the RC ($\tau_{RC}$). We assume $\tau_B$/$\tau_{RC}$ = $D_{ratio}$, in the basic variant of simulation $D_{ratio}$=600.
Considering the energetic landscape for the RC motion, it is described by the potential U, which has two components: the energetic barrier separating the closed and open states ($U_{TP}$) in form of a single peak from TP-1 to TP+1, and the linear drift force component (A) representing the channel activation level. (When the open and closed states are equally probable, A = 0. If the open states are preferred, A<0. If the closed states are preferably occupied, A>0.)
At each time step of the simulation $\tau_{RC}$, the RC jumps to the left or right, and the corresponding probabilities $p$ and $q$ are given by:
\begin{equation}
    p = \frac{1}{2} - \frac{\Delta U}{4kT}
\end{equation}
and
\begin{equation}
    q = \frac{1}{2} + \frac{\Delta U}{4kT}
\end{equation}
respectively \cite{berg1993random}.

As a result of model simulation, one obtains the series of ones and zeros which correspond to the conducting and non-conducting states of the channel recognized in the subsequent reaction coordinate time units, respectively.
Based on such a series, the corresponding sequences of dwell times of the succeeding functional states can be easily constructed.
In this work, we present the results of a model simulation based on 5 independent repetitions, each of them last 10000000 reaction coordinate time units.  

\section{Results}

\subsection{Effects of gating machinery}

The Hurst exponents obtained for the experimental data for different potassium channel types are presented in Table~\ref{Tab:Hurst_exponents}. All examined dwell-time series exhibit the trend-reinforcing behavior ($H_{RS}>$0.5).
As one can see, the compared channels substantially differ by a single channel conductance (G).
MitoTASK-3 channel is a weakly-rectifying channel. Thus, two different regimes of channel conductance have been observed at membrane hyper- and depolarization.
In general, the smaller memory effect can be observed for the mitoKv1.3, mitoTASK-3 and TREK-2-like channels, that have low single-channel conductance among the analyzed channel types. Whereas the highest $H_{RS}$ values correspond to the large-conductance BK channels from the glioblastoma cells.

An interesting relation between $G$ and $H_{RS}$ can be observed for the mitoTASK-3 in the $G_1$ = 13.5 [pS] and  $G_2$ = 36.0 [pS] conductance regimes.
These results seem to contradict the initial hypothesis on the relation between channel's conductance and the magnitude of the Hurst effect. However, the conductance rather sets an upper limit for $H_{RS}$, and does not directly translate to an exact value of $H_{RS}$. 
The observed phenomenon can be explained within the considered framework of dwell time fluctuations after assuming a model of the rectification performed by the channel. We have chosen to consider a model where unstable open conformations are (partially) removed from the time series - this lowers the average conductance (for peaks beyond sampling rate) and lifts $H_{RS}$ due to general removal of short open dwell-times, which were affected by the interactions with SF. We could successfully model this process (see below).

\begin{table}
\caption{The mean values of Hurst exponents calculated by the R/S ($H_{RS}$) and DFA ($H_{DFA}$) methods for dwell-time series of different potassium channels. The uncertainity is given by the standard error. $G$ denotes channel conductance.\\} 
\label{Tab:Hurst_exponents}
\centering
\begin{tabular}{p{0.55\textwidth}p{0.15\textwidth}p{0.15\textwidth}p{0.15\textwidth}}
\hline
\textbf{Channel type} & \textbf{G [pS]} & \textbf{$H_{RS}$} & \textbf{$H_{DFA}$}\\
\hline
mitoTASK-3 from HaCaT keratinocytes, U = 90 mV & 13.5$\pm$2.7 & 0.61$\pm$0.02 & 0.78$\pm$0.01 \\
\hline
mitoTASK-3 from HaCaT keratinocytes, U = -90 mV & 36.0$\pm$1.0 & 0.58$\pm$0.02 & 0.75$\pm$0.05  \\
\hline
mitoKv1.3 channel from gerbil hippocampus & 107.4$\pm$3.2 & 0.57$\pm$0.01 & 0.63 $\pm$0.05  \\
\hline
TREK-2-like channels from rat pyramidal neurons & 126.7$\pm$8.2 & 0.60$\pm$0.02 & 0.66 $\pm$0.05  \\
\hline
BK from glioblastoma (U87-MG), U = 40 mV & 236.8$\pm$8.4 & 0.75$\pm$0.07 & 0.81$\pm$0.07\\
\hline
\end{tabular}
\end{table}

To investigate quantitatively our hypotheses regarding the impact of SF fluctuations (and therefore of the conductance) on the Hurst exponent, we simulated the effect of such perturbations on a dwell-time series with strong correlations (glioblastoma U87-MG at 20mV), where one of the traces revealed $H=0.81 \pm 0.04$. 

\begin{figure}[H]
\centering
\includegraphics[width=1.0\linewidth]{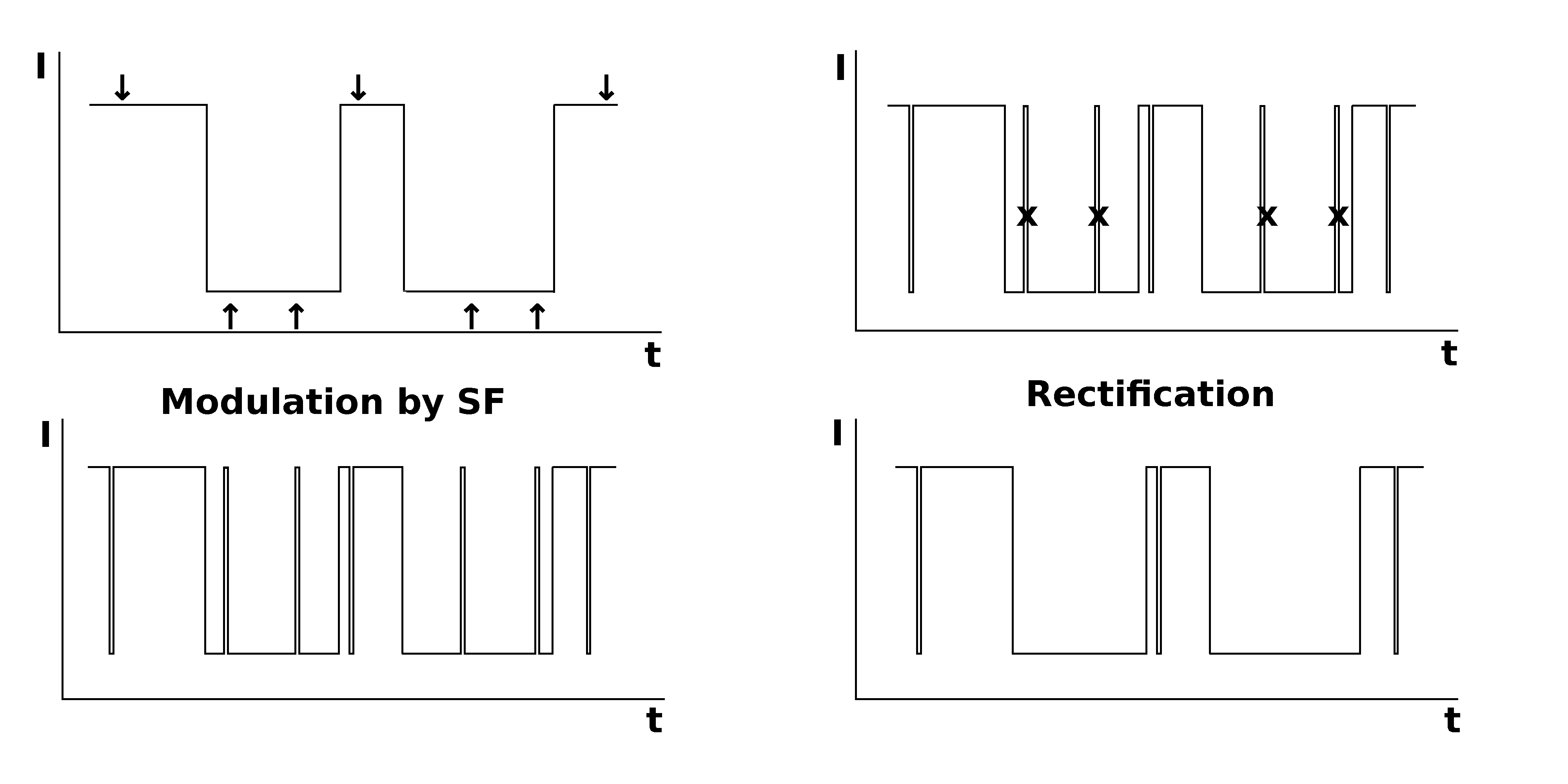}
\caption{Illustration of the modulation imposed on the ionic current of the channel by fast switching of the selectivity filter (left panel) and the idea of rectification, which reduces permeability to the ions and increases the Hurst exponent by eliminating to some extent the modulation imposed by SF (right panel).}
\label{fig:modeling}
\end{figure}

In such dwell-time series we have introduced rapid conductance-reducing fluctuations, generated by SF (which occasionally are expected to be long enough to be caught by the sampling device). So with probability equal to one event per $0.3ms$ we introduced peaks of duration $0.1ms$ of opposite conformation (see the left panel of Fig. \ref{fig:modeling}).

This procedure destroyed the Hurst effect, revealing $H=0.57 \pm 0.01$. What's more interesting, the effect related to the prolongation of initial ,,low'' RS slope at cost of the ,,high'' RS slope, because the result of the described procedure was gradually canceled when length of the dwell-time sample was increased (e.g. $H=0.63 \pm 0.01$ for $n=8192$, $H=0.74 \pm 0.03$ for $n=16384$), and the ,,high'' slope was not entirely destroyed in such case. However, samples of length larger than about $2000$ events are rarely collected.

The above procedure was also challenged with the results of the TASK channels, which display rectification. In such channels, the conductance depends on orientation, and so does the Hurst exponent: it is larger for the orientation of lower conductance. We have modeled rectification of lower conductance by removal of unstable opening events (i.e., events that take less than 0.7ms of time - see the right panel in Fig. \ref{fig:modeling}). This models how the rectification ,,undoes'' the changes introduced by fluctuations of the SF, and we were able to obtain an increase in $H$ from $0.57 \pm 0.01$ to $0.61 \pm 0.02$.

\subsection{Effects of channel activation}

\begin{figure}[ht]
    \centering
    \includegraphics[width=0.99\textwidth]{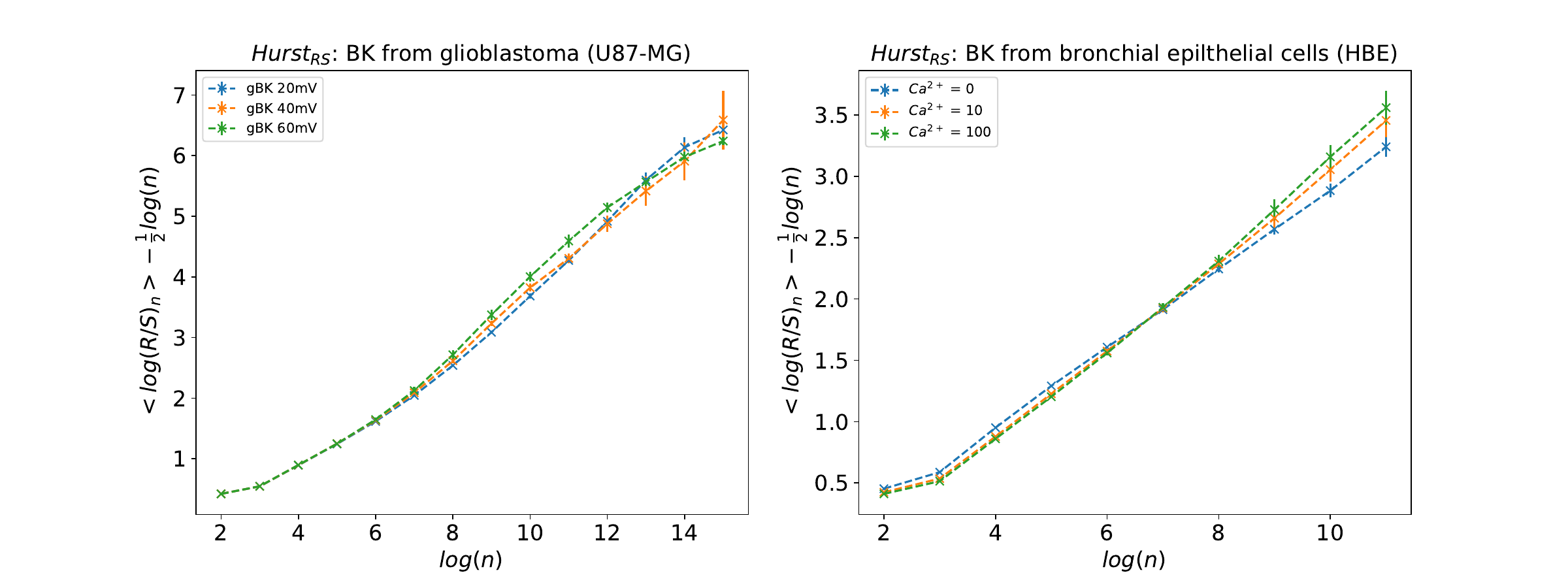}
    \caption{The average values of log(R/S) in the function of log(n) calculated for the scaling range of the full sequences: $log(n) \subset [2,15]$. The half of the values of each point on the abscissa axis have been subtracted from the corresponding values on the ordinate axis for the better visualization of the differences (which corresponds to the trend line at $H_{RS}$=0.5). In such framework, the uncorrelated R/S plot is a flat line, persistent R/S plot is a growing curve, antipersistent R/S plot is a decreasing curve. Each point was presented together with the standard error of the mean. The left panel characterizes the scaling functions for the BK channels from glioblastoma at different values of membrane potential. The right plot summarizes the impact of calcium concentration on the activity of BK channels from bronchial epithelial cells.}
    \label{fig:scalling_activation}
\end{figure}

\begin{figure}
    \includegraphics[width=.5\textwidth]{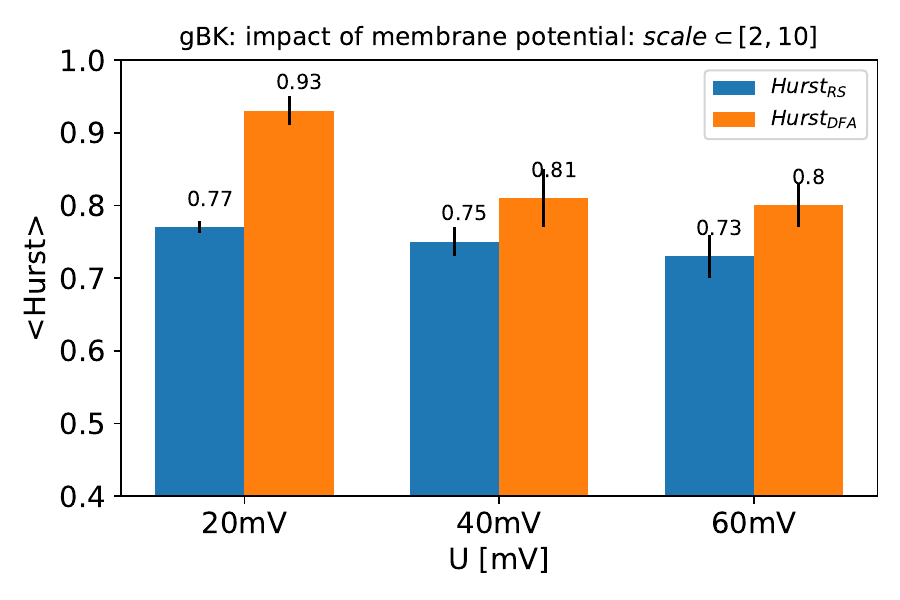}
    \includegraphics[width=.5\textwidth]{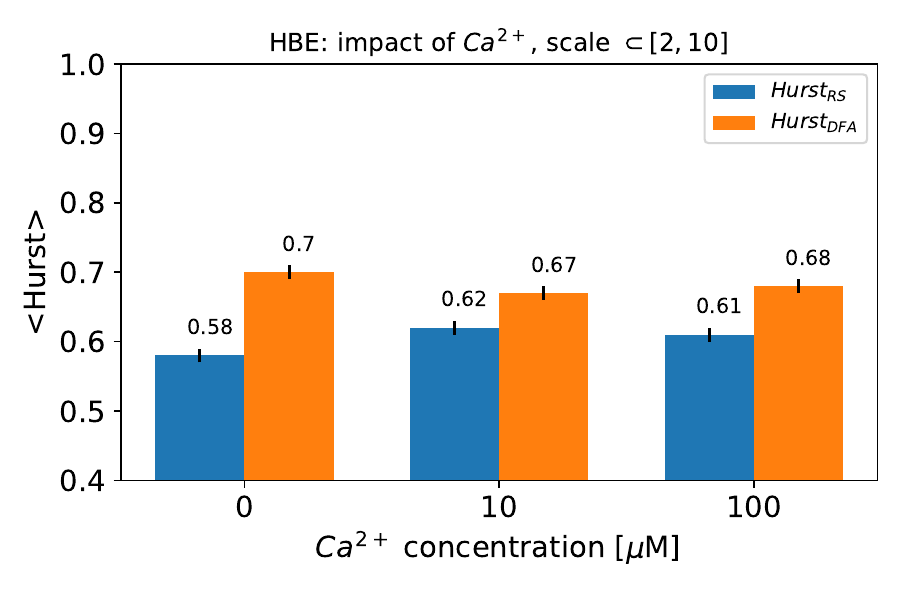}
     \includegraphics[width=.5\textwidth]{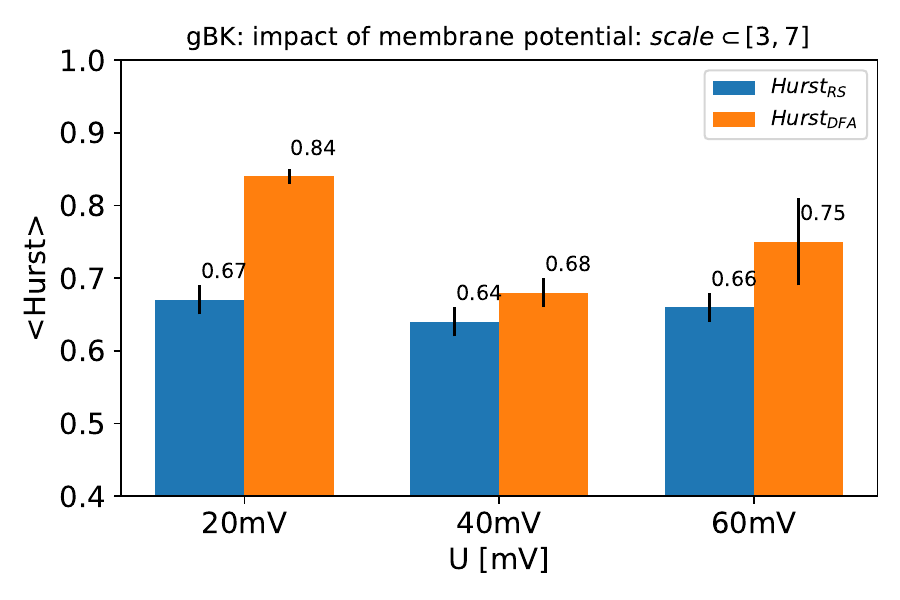}
    \includegraphics[width=.5\textwidth]{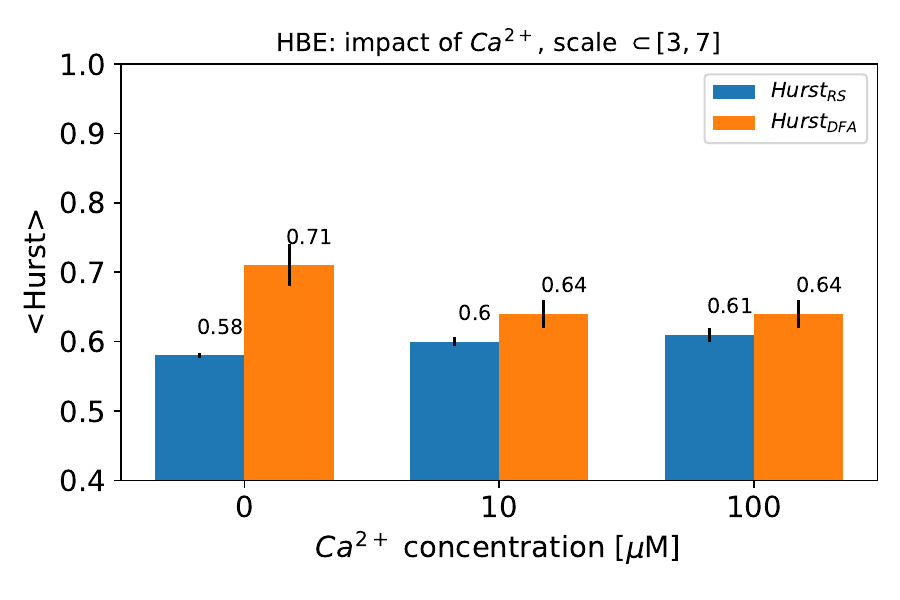}
     \includegraphics[width=.5\textwidth]{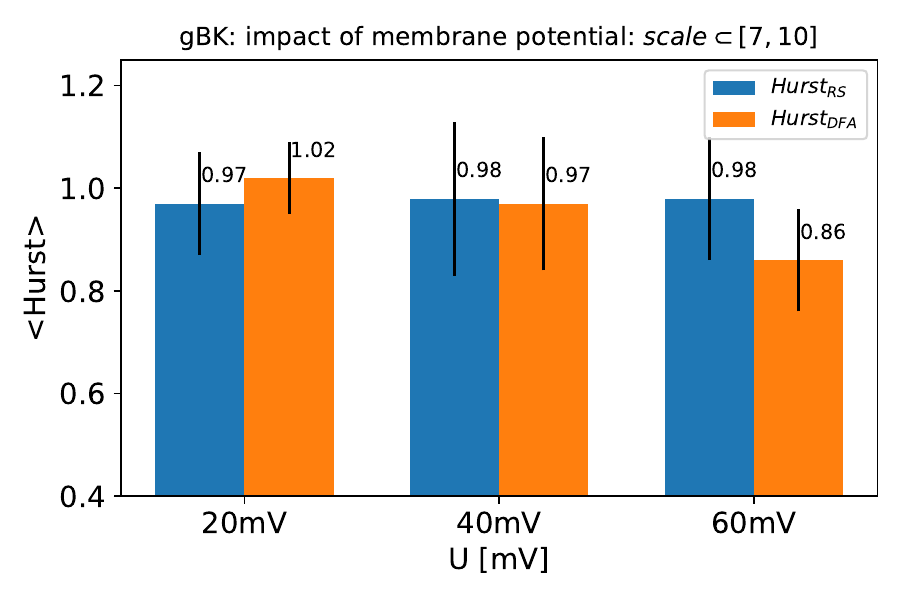}
    \includegraphics[width=.5\textwidth]{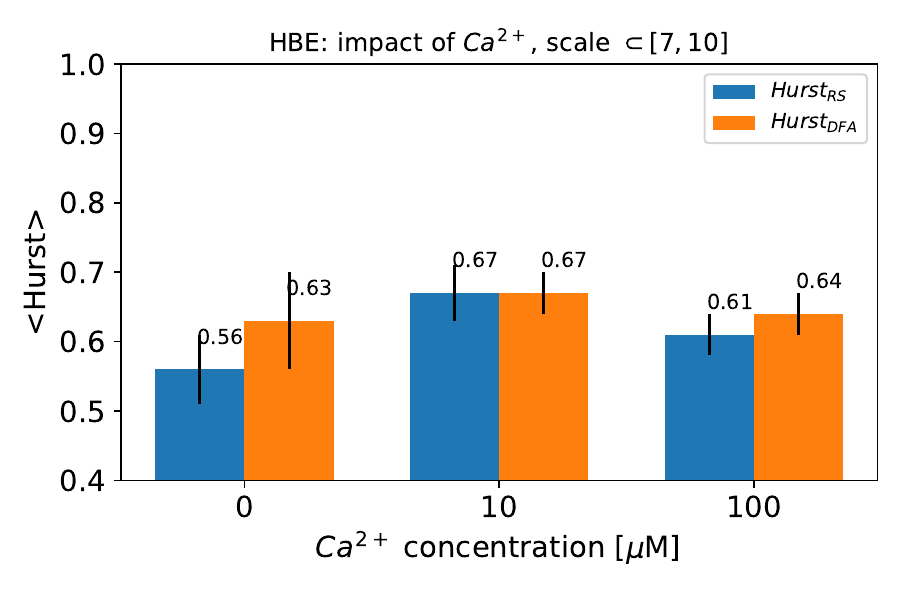}
    \caption{The average values of Hurst exponent calculated by the DFA and R/S technique at the selected scaling range for the BK channel activation by voltage and $Ca^{2+}$. The upper panels characterize the  scale $\subset$ [2,10]. The middle panels are assigned to the small scaling range: scale $\subset$ [3,7]. The lower figures represent the large scaling range: $\subset$ [7,10].}
    \label{fig:Hurst_exponents_UCa_activation}
\end{figure}

The effects of channel activation on the Hurst memory are relatively weak and frequently non-monotonic, as could be observed in the literature (Table \ref{Tab:Hurst_literature}).
Here, this impact is considered for the BK channels from the human glioblastoma cells for both generic factors that increase its open state probability; i.e., membrane potential (U) and the $Ca^{2+}$ concentration on the cytosolic side of membrane (Table \ref{Tab:Hurst_exponents_activation}, Figs. \ref{fig:scalling_activation}, \ref{fig:Hurst_exponents_UCa_activation}).
According to the former observations and the results provided in this work, it seems that the coupling of regulatory domains (like voltage-sensing domain or ligand-sensing domain) at different levels of their activation with the channel gate introduces neither strong random patterns nor highly oriented ones to the switching machinery.
However, such an effect certainly is possible, as the cumulative action of calcium and voltage activation, and therefore the movement of relevant protein domains, has a potency to considerably change the value of the Hurst exponent; $H_{RS}$ by about $0.07$ and $H_{DFA}$ by about $0.11$ (Table \ref{Tab:Hurst_exponents_activation}).

\begin{table}
\caption{The mean values of Hurst exponents calculated by the R/S ($H_{RS}$) and DFA ($H_{DFA}$) methods for dwell-time series of the BK channels states obtained at different levels of channel activation by membrane depolarization and $Ca^{2+}$ concentration. The uncertainity is given by the standard error. $G$ denotes channel conductance, $p_{op}$ is the open state probability.\\} 
\label{Tab:Hurst_exponents_activation}
\centering
\begin{tabular}{p{0.4\textwidth}p{0.15\textwidth}p{0.15\textwidth}p{0.15\textwidth}p{0.15\textwidth}}
\hline
\textbf{Channel type} & \textbf{$p_{op}$} & \textbf{$G [pS]$} &\textbf{$H_{RS}$} & \textbf{$H_{DFA}$}\\
\hline
BK from glioblastoma (U87-MG), U = 60 mV & 0.88$\pm$0.02 & 224.1$\pm$6.9 & 0.73$\pm$0.02 & 0.80$\pm$0.07 \\
BK from glioblastoma (U87-MG), U = 40 mV & 0.80$\pm$0.02 & 236.8$\pm$8.4 & 0.75$\pm$0.07 & 0.81$\pm$0.07  \\
BK from glioblastoma (U87-MG), U = 20 mV & 0.74$\pm$0.05 & 239.6$\pm$33.0 & 0.77$\pm$0.02 & 0.93$\pm$0.03  \\
\hline
BK from bronchial epithelial cells (HBE), $Ca^{2+}$ = 0 $\mu$M & 0.30$\pm$0.06 & 224.4$\pm$3.9 & 0.58$\pm$0.01 & 0.70 $\pm$0.02  \\
BK from bronchial epithelial cells (HBE), $Ca^{2+}$ = 10 $\mu$M & 0.83$\pm$0.03 & 208.6$\pm$8.0 & 0.62$\pm$0.01 & 0.67 $\pm$0.02  \\
BK from bronchial epithelial cells (HBE), $Ca^{2+}$ = 100 $\mu$M & 0.91$\pm$0.01 & 213.1$\pm$6.5 & 0.61$\pm$0.01 &  0.68 $\pm$0.08  \\
\hline
\end{tabular}
\end{table}

The plots of the rescaled range vs. the window size (Fig. \ref{fig:scalling_activation}), revealed that the variability between different activation levels both by membrane depolarization as well as the increase of $Ca^{2+}$ concentration is more evident in the long subseries regime (for log(n) $\geq$ 7) than for the short subseries regime (for log(n) $\leq$ 7).
Therefore, in Fig. \ref{fig:Hurst_exponents_UCa_activation} the obtained Hurst exponents are presented in three variants; i.e., in the broad range of log(n), for short subseries (where log(n) $\subset$ [3,7]) and for long subseries (where log(n) $\subset$ [7,10]).
Particularly, in terms of the voltage-activation of the BK channels in glioblastoma cells, one can observe two different values of scaling for long and short window sizes (log(n)). This kind of behavior has been already reported in literature before \cite{kochetkov1999non, siwy2001application, siwy2002correlation}.

What is also worth noticing, it is that for the BK channels from glioblastoma we had extraordinary long dwell-time series at our disposal. They described at least tens of thousands acts of switching between open and closed states. Such kind of data allowed us to track the rescaled range scaling function up to the relatively large scales.
It turns out, that the power-law scaling is conserved in the full analyzed range (up to the $2^{15}$-long windows).

The results above reveal also that memory effect of the BK channels in bronchial epithelial cells is relatively small, no matter its high conductance. This will be broadly commented in the Discussion.

\subsection{Effects of auxiliary regulating subunits, membrane type and different channel isoforms}

\begin{figure}[h]
    \centering
    \includegraphics[width=0.99\textwidth]{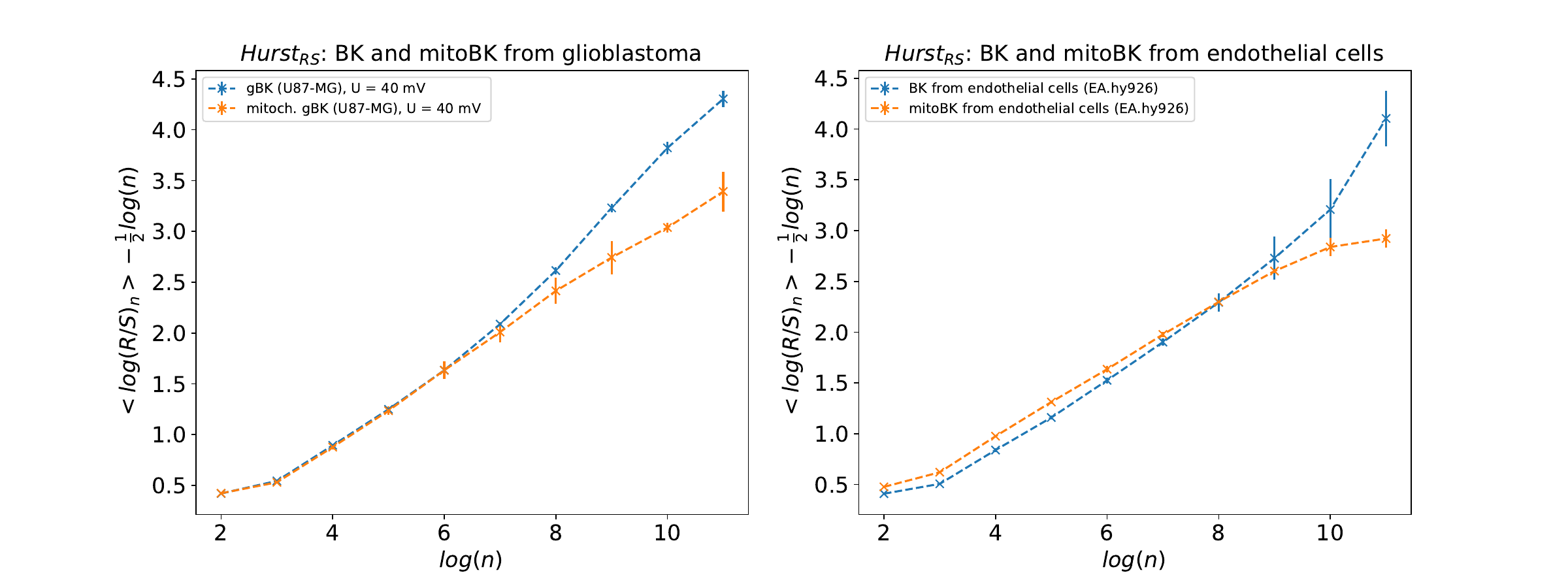}
    \caption{ The average values of log(R/S) in the function of log(n) calculated for the scaling range
of the full sequences: s $\in$ [2, 11]. The half of the values of each point on the abscissa axis have been
subtracted from the corresponding values on the ordinate axis for the better visualization of the
differences. Each point was presented together with the standard error of the mean. The left panel
characterizes the scaling functions for the glioblastoma cell membrane channels and their mitochondrial equivalents.
The right plot sets together the mitochondrial and plasma membrane- BK channels of endothelial cells. }
    \label{fig:scalling_cell_mito}
\end{figure}

\begin{figure}[H]
    \centering
    
    \includegraphics[width=0.49\textwidth]{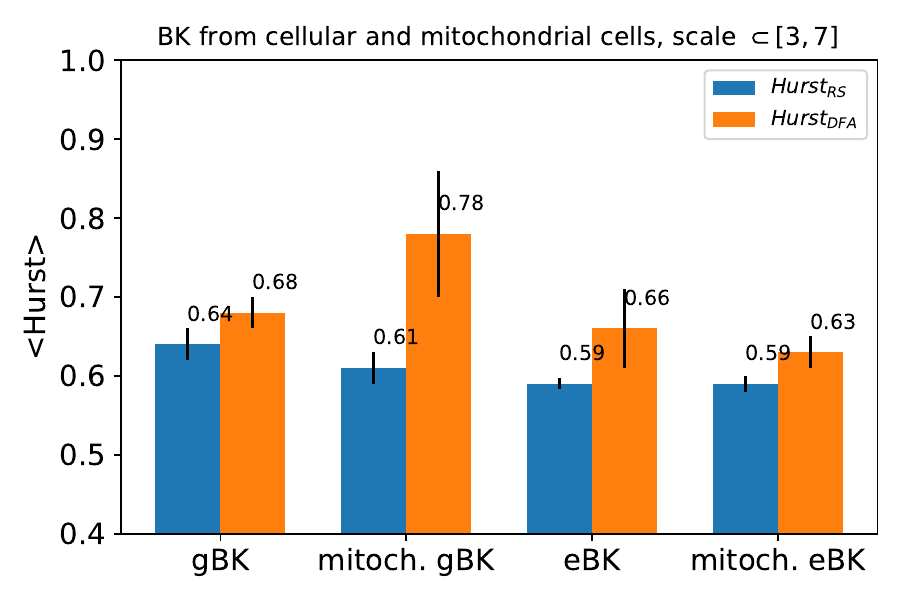}
    \includegraphics[width=0.49\textwidth]{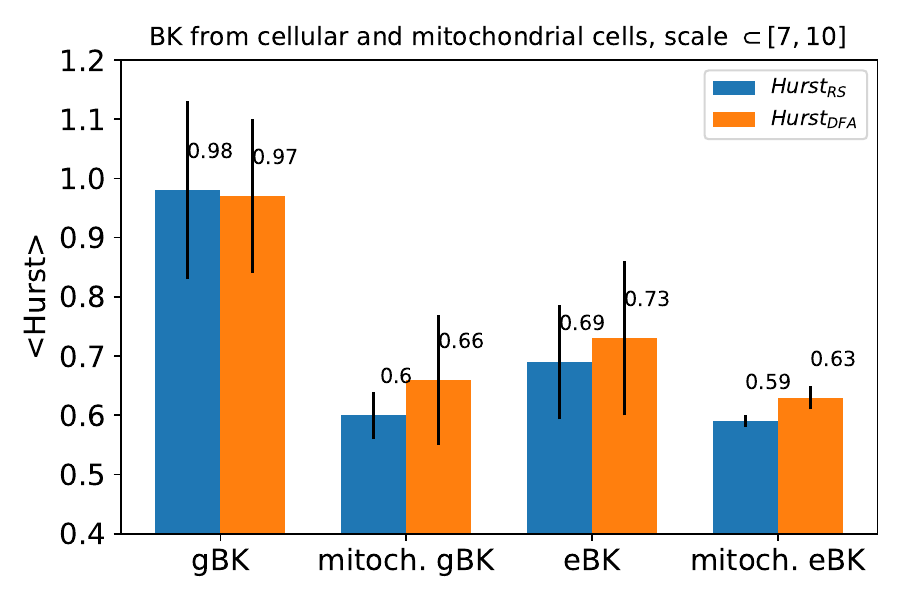}
    \includegraphics[width=0.5\textwidth]{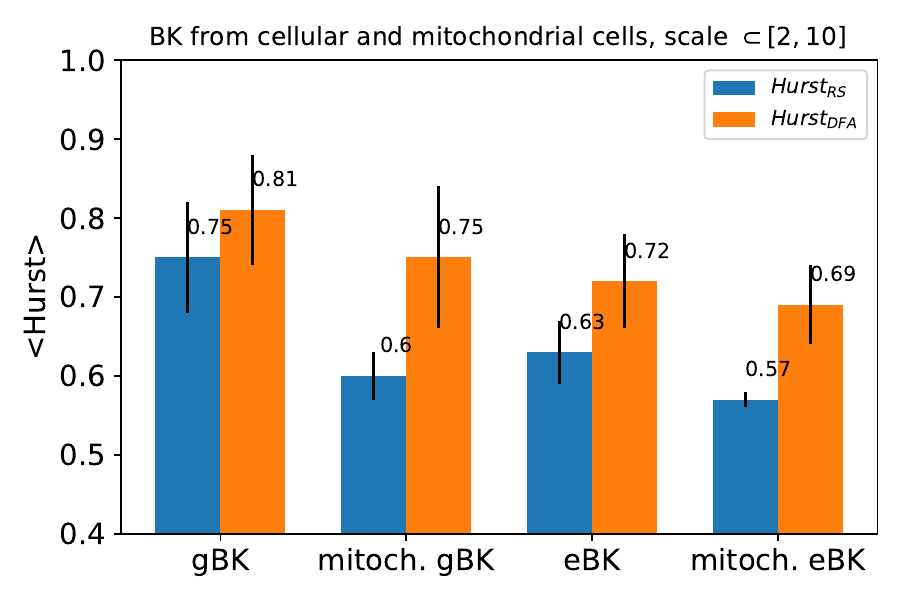}
    \caption{The average values of Hurst exponent calculated by the DFA and R/S technique at the selected scaling ranges for the plasma membrane- and mitochondrial BK channels from glioblastoma and endothelial cells. First subplot characterizes the scale $\subset$ [2,10]. Second subplot is assigned to the small regime of n: scale $\subset$ [3,7]. The last subplot represents the large windows: $\subset$ [7,10].}
    \label{fig:Hursts_cell_mito_chart}
\end{figure}

The next comparison of the memory effect we would like to present is the juxtaposition of BK channels from different cell types.
As a running example we take the activity of the BK channel in glioblastoma cells and compare their Hurst exponents with those found in the endothelial cells.
Additionally, we present the $H$ exponents for the mitochondrial variants of these channels. The results are presented in Table \ref{Tab:Hurst_exponents_mitocell} and Figs. \ref{fig:scalling_cell_mito}, \ref{fig:Hursts_cell_mito_chart}.

\begin{table}[H]
\caption{The mean values of Hurst exponents calculated by the R/S ($H_{RS}$) and DFA ($H_{DFA}$) methods for dwell-time series of plasma membrane- and mitochondrial BK channels from two different cell types (at U = 40 mV and calcium concentrations ensuring comparable activation levels). The uncertainty is given by the standard error. $G$ denotes channel conductance. For both cell lines, the leading type of channel auxiliary regulating subunit is given, for which the highest expression level was exhibited in original study \cite{bednarczyk2013large, bednarczyk2013putative}.\\} 
\label{Tab:Hurst_exponents_mitocell}
\centering
\begin{tabular}{p{0.4\textwidth}p{0.15\textwidth}p{0.15\textwidth}p{0.15\textwidth}p{0.15\textwidth}}
\hline
\textbf{Channel type}	& \textbf{Regulating subunits}& \textbf{G [pS]} & \textbf{$H_{RS}$} & \textbf{$H_{DFA}$}\\
\hline
BK from glioblastoma (U87-MG) & $\beta$4 & 236.8$\pm$8.4 & 0.75$\pm$0.07 & 0.81$\pm$0.07  \\
mitoBK from glioblastoma (U87-MG) & $\beta$4 & 276.7$\pm$28.9 & 0.60$\pm$0.03 &  0.75$\pm$0.09   \\
\hline
BK from endothelial cells (EA.hy926) & $\beta$2 & 241.3$\pm$10.6 & 0.63$\pm$0.04 &  0.72$\pm$0.06 \\
mitoBK from endothelial cells (EA.hy926) & $\beta$2 & 266.5$\pm$0.6 & 0.57$\pm$0.01 &  0.63$\pm$0.05  \\
\hline

\end{tabular}
\end{table}

According to the presented results, the memory effects in BK channels from endothelial cells were smaller in comparison with the ones from glioblastoma.
The differences between the cell types are higher than the differences in H for different levels of channel activation (as presented in the previous paragraph).
They are even comparable with the impact of gating machinery, as shown for the different channel types (Table \ref{Tab:Hurst_exponents}).

Our results also showed that the mitoBK channels exhibit lower memory effects than their analogs from the plasma membrane for both analyzed cell types (Tab. \ref{Tab:Hurst_exponents_mitocell}, Figs. \ref{fig:scalling_cell_mito}, \ref{fig:Hursts_cell_mito_chart}).
The differences between their scaling behavior are mostly visible for large scales (see Fig. \ref{fig:scalling_cell_mito} from log(n) = $2^7$ to log(n) = $2^{11}$.

\begin{figure}[H]
    \includegraphics[width=.5\textwidth]{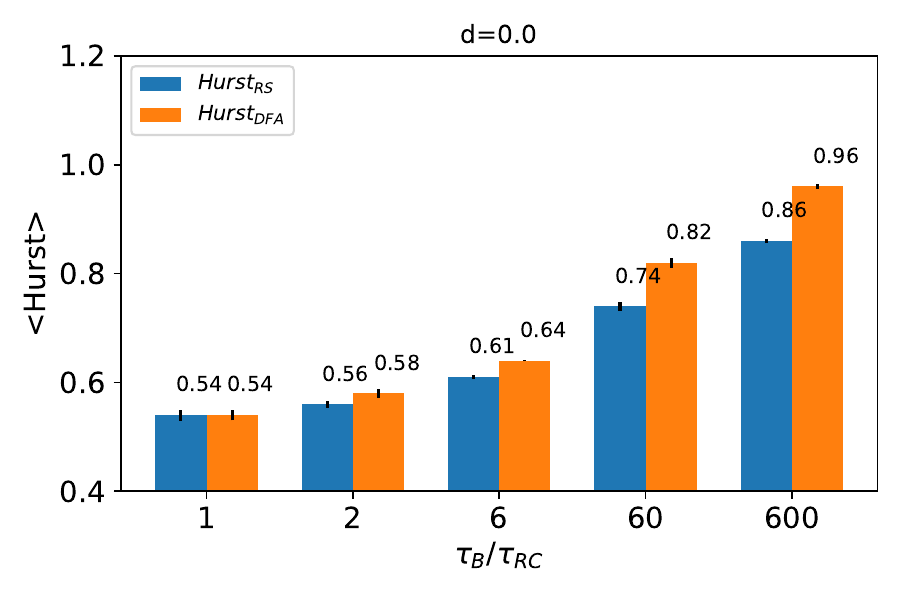}
    \includegraphics[width=.5\textwidth]{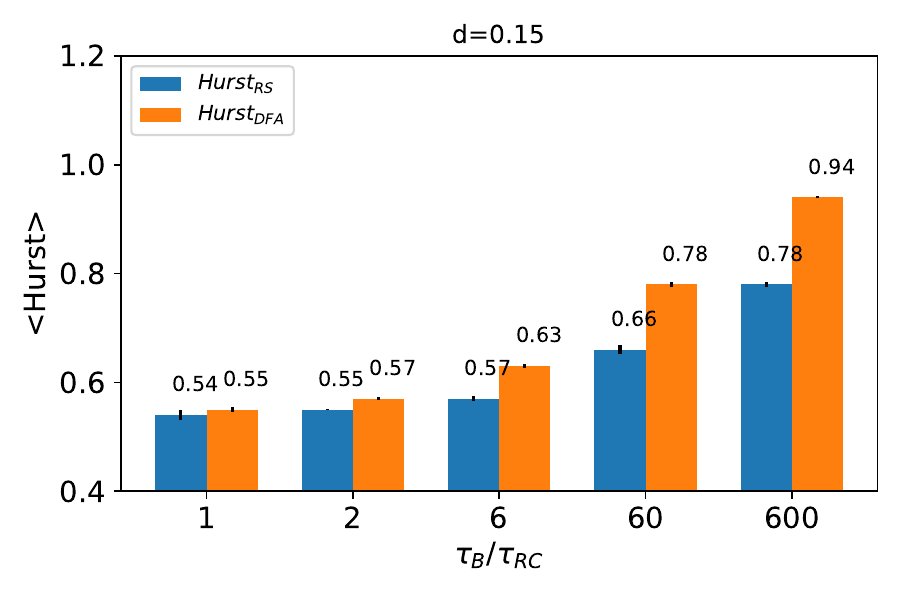}
    \caption{The values of Hurst exponent evaluated by the R/S and DFA algorithms for the dwell-time series obtained by the simulation of model of channel gate proposed in \cite{wawrzkiewicz2012simple} at different values of the $\tau_{B}/\tau{RC}$ parameter at drift force d=0 (open and closed conformations are equally probable) and d=0.15 (open channel conformations are preferred).}
    \label{fig:Hurst_exponents_model_2012}
\end{figure}

The differences between the mitochondrial and cell membrane- variants of channels prompted us to inspect the effects exerted by the channel surroundings in a more detailed way.
Consequently, we have simulated the random walk model of a channel gate \cite{wawrzkiewicz2012simple} that considers the impact of the fluctuation of the membrane where the channel is immersed.
Due to the differences in the environment that the channels operate in mitochondria and plasma membrane, including membrane biochemistry and biophysics, it was reasonable to simulate gating for different values of the parameter that regulates the impact of channel surroundings on its gate, i.e., the $D_{ratio} = \tau_B/\tau_{RC}$, which is the ratio of time scales of boundaries and reaction coordinate movement (or - in other words - of the diffusion coefficients, of membrane fluctuations and gating machinery elements fluctuations, which depend strongly on viscosity, that may be a factor here due to membrane composition differences).
The results of the Hurst analysis for the obtained simulated dwell-time series are presented in Fig. \ref{fig:Hurst_exponents_model_2012}.

Our simulation model relies on the speed of fluctuations of the membrane, which interacts with channel's protein.
The smaller (or just more prone to thermal motion) is the fluctuating object, the smaller the amplitude of fluctuations, the faster they are. In those terms the Hurst effect reduces (Fig. \ref{fig:Hurst_exponents_model_2012}).
We are convinced that similar effects may occur in the BK/mitoBK system, due to the differences between the mitochondrial and plasma membrane.

\section{Discussion} 

The molecular origin of the Hurst memory effect in the experimental traces describing ion channel activity needs a mechanistic explanation.
For this sake, in this work we decided to address that question by the analysis of the experimental data (for various channel types and their isoforms at different external conditions) as well as the performed the appropriate simulations.

According to the obtained results, memory seem to be an universal feature of the potassium channels’ gating, but different factors can affect (improve or destroy) the actual strength of the long-range correlations.
Among them, the allosteric AG-SF coupling, which determines the single-channel conductance by stronger or weaker stabilization of open states, is expected to shape the attainable strength of the long-range correlations. The large-conductance channels (with stable, rarely perturbed open state conformations) should be prone to exhibit the trend-reinforcing behavior, the small-conductance channels (with less stable, frequently perturbed open state conformations) should be related to weaker memory effects (Table. \ref{Tab:Hurst_exponents}). This relation can be however distorted by the additional interfering processes.
One of such processes is the channel rectification, as was revealed for the mitoTASK-3 channels (Table \ref{Tab:Hurst_exponents}) and the corresponding simulations (according to the idea presented in Figure \ref{fig:modeling}).

Another factor that deeply bias the gating dynamics, and consequently the long-range memory effect, are the small mechanistic differences between channel isoforms and the possible inactivation by the auxiliary regulating subunits (like for the BK channels coordinated with channel-inactivating auxiliary subunits, Table \ref{Tab:Hurst_exponents_mitocell} and Figure \ref{fig:Hursts_cell_mito_chart}).
BK channels are encoded in humans by the Kcnma1 gene wherever they are expressed. However, the BK exonic composition in different locations can vary due to the alternative splicing during the Kcnma1 gene transcription. This leads to a tissue-specific functional heterogeneity within the BK channels (including such features as functional characteristics including voltage/$Ca^{2+}$ sensitivities, response to phosphorylation, and subcellular localization)
\cite{latorre2017molecular}.
It turns out that in glioblastoma cells, BK channels are expressed in specific isoforms which differ from the other exons by the enhanced sensitivity to cytosolic $Ca^{2+}$ concentration \cite{ransom2002bk}.

The phenotypic diversity of BK channels can be also enhanced by the possibility of associating the channel protein with different types of regulatory $\beta$ (1-4) and $\gamma$ (1-4)
subunits \cite{latorre2017molecular, balderas2015mitochondrial}. 
In the case of the BK channels in the glioblastoma, they form complexes with the $\beta$4 subunits \cite{bednarczyk2013putative}, while their analogs in the endothelium are coordinated with the $\beta$2 subunits \cite{bednarczyk2013large}.
Because the auxiliary regulating subunits ($\beta$ subunits) are responsible for the regulation of channel sensitivity to $Ca^{2+}$ and the channel proneness to inactivation \cite{contreras2012modulation, balderas2015mitochondrial}, they are expected to affect the dwell-time patterns in gating dynamics, and consequently the $H$ values.
This could be observed in our results as lowering the trend-reinforcing effect in the case of endothelial channels (where the channels form mainly the $\alpha$--$\beta$2 complexes) in comparison to the ones from glioblastoma (where the channels form mainly the $\alpha$--$\beta$4 complexes), see Table \ref{Tab:Hurst_exponents_mitocell} and Figure \ref{fig:Hursts_cell_mito_chart}. 
It should also be mentioned, that the $\beta$2 subunits can distinctively hamper BK channel dynamics even by induction of inactivated states, thus, they are anticipated to more profoundly modulate the native dwell-time distributions than the other types of $\beta$ subunits \cite{orio2005differential}.
The lowering of the memory effect by the $\beta$ subunits could be also observed by the $Ca^{2+}$ dependency of $H$ shown in Table \ref{Tab:Hurst_exponents_activation} because the BK channels from the HBE cells can be coordinated to the auxiliary subunits of $\beta$3 and $\beta$4 types \cite{sek2021identification} and some $\beta$3 variants can induce a rapid channel inactivation \cite{li2016modulation}, that ought to disturb the correlated patterns in channel activity.
From a mechanistic point of view, the $beta$ subunits are expected to modulate the dwell times in a similar way as SF, but rather on a larger time scale (so possibly for larger time windows on the R/S plot). Please note, that such interaction with beta subunits was not considered in the initial proposal of Hurst exponent-conductance relation, and therefore it does not contradict it. It merely provides a way to bring $H$ down even if the SF interactions don't imply this.

In this study we also showed that the mitochondrial BK channels exhibit lower memory effects than their plasma membrane analogs for both glioblastoma and endothelial cells (Tab. \ref{Tab:Hurst_exponents_mitocell}, Figs. \ref{fig:scalling_cell_mito}, \ref{fig:Hursts_cell_mito_chart}). 
Considering the possible reasons for the mentioned differences in $H_{RS}$ and $H_{DFA}$ between the plasma membrane- and mitochondrial variants of the BK channels, it can be emphasized that the mitoBK channels are mainly expressed when the Kcnma1 undergoes splicing to the DEC isoform \cite{singh2013mitobkca} (or, likely, some other mitochondria-specific exons).
The DEC splice variant is recognized by the presence of an insertion of a 50-amino acid C-terminal sequence which prevents expression of the BK channel in the plasma membrane.
It is possible, yet still not clear, whether the BK-DEC splice variant imposes functional changes in the channel's activity in relation to the other BK channel isoforms found in the cell membrane.

Another factor that can result in different characteristics of mitoBK and BK channel gating is the discrepancies between the biophysical and biochemical properties between the plasma membrane and the mitochondrial membranes. In contrast to plasma membranes, mitochondrial membranes contain high levels of cardiolipin (ca. 15\% of total phospholipids), while the levels of sphingolipids and cholesterol are relatively low \cite{schenkel2014formation}.
The different compositions of cellular and mitochondrial membranes can result in different viscous properties and changes in lipid-protein and protein-protein interactions that affect gating dynamics \cite{lee2004lipids, brown2017membrane, tillman2003effects, duncan2017protein, cordero2018lipids}.
The importance of the channel surroundings characteristics for the Hurst memory effect, which can contribute to the differences between the mitoBK/BK channels, has been confirmed by the results of model simulation (Fig. \ref{fig:Hurst_exponents_model_2012}).

The results obtained here may improve our understanding of shaping the channel activity patterns by different component processes (selectivity filter fluctuations, its coupling with the activation gate, interaction with regulatory subunits, rectification etc.).
The interplay between the Hurst effect and the allosteric regulation of ion channels by different factors holds significant implications for the field of electrophysiology, because it gives a new light for the processes occurring at a relatively large time scale.
Thus, performing the electrophysiological measurements at a longer time than usual (even for tens to hundreds of seconds) can introduce new interesting information. For example, different characteristics of the BK channels in glioblastoma compared to other cell types have been unraveled for the large time scales (Figure \ref{fig:scalling_activation}).

The important limitation of this study is the number of channel types analyzed. 
We inspected the memory effect for the wider range of potassium channels than ever before (BK, mitoBK, mitoTASK-3, mitoKv1.3, TREK-2-like channels), to get a possibly detailed picture of this phenomenon. But still these are only four independent channel types from the widest channel family, which has tens of members.
We have carried out the Hurst analysis for the simulated data to explain some aspects of gating-memory relations. One more time, in that context we also used limited model approaches (according to the simplified approach to rectification, Figure \ref{fig:modeling} or our model \cite{wawrzkiewicz2012simple} proposed previously).
From this perspective, examination of the more simulated data based on different assumptions would be also interesting.

\section{Conclusions}
We made a new insight into the problem of dwell time correlations in ion channels. All potassium channels analyzed in this work exhibited the log-range memory effect. We pointed classes of channels, which may display larger Hurst exponents than the others and explained the reasons for this. The results are supported by current molecular dynamics models of the SF operation \cite{kopec2019molecular, mironenko2021persistent}.

The channel conductance should set an upper limit for the Hurst effect. Namely, the large-conductance channels are supposed to exhibit the trend-reinforcing behavior (their open states are highly stabilized). Whereas, the small-conductance are anticipated to show weaker memory effects. Nevertheless, this relation can be easily disturbed by the additional interfering processes, including the channel rectification (like for the mitoTASK-3 channels) or its inactivation by the auxiliary regulating subunits (like for the BK channels coordinated with $\beta$2 and $\beta$3 subunits).
Another factor that deeply bias the gating dynamics, and consequently the Hurst effect, are the interactions of the channel protein with its surroundings (e.g., lipid microenvironment) and the presence of different amino acid inserts in particular channel isoforms, as can be seen when the $H$ exponents are compared for the BK channels from mitochondrial and plasma membranes.

Of course, the hypotheses proposed here demand further validation by more experimental recordings between comparable channels.
We used all of the experimental data, that we had, searched out the literature, and performed the simulation of a new, fruitful model of channel rectification, as well as revisited the old one \cite{wawrzkiewicz2012simple}, to explain different aspects of the long-range correlations within the dwell-time series describing the ion channel activity.
The results throw some new light of the memory effect in channel gating. Nevertheless, they are still not fully satisfactory, especially due to a limited amount and length of the data.
Hopefully, an increased interest in the research on the relations between channel conductances, rectification, and interactions with auxiliary domains on the Hurst effect will follow.

\section*{Acknowledgments}

This research was supported in part by the grant \textit{Miniatura 6}
no. 2022/06 /X/ST6/01133 from the National Science Centre (NCN, Poland) to P.T. and the 
System of Financial Support for Scientists and Research Teams – SGGW No. S23007 to P.B. (Piotr Bednarczyk).

\section*{Conflict of interest}

All authors declare no conflicts of interests.

\appendix
\section{}\label{Appendix A}
The samples of representative patch-recordings analyzed in current research.

\begin{figure}[H]
    \includegraphics[width=0.9\textwidth]{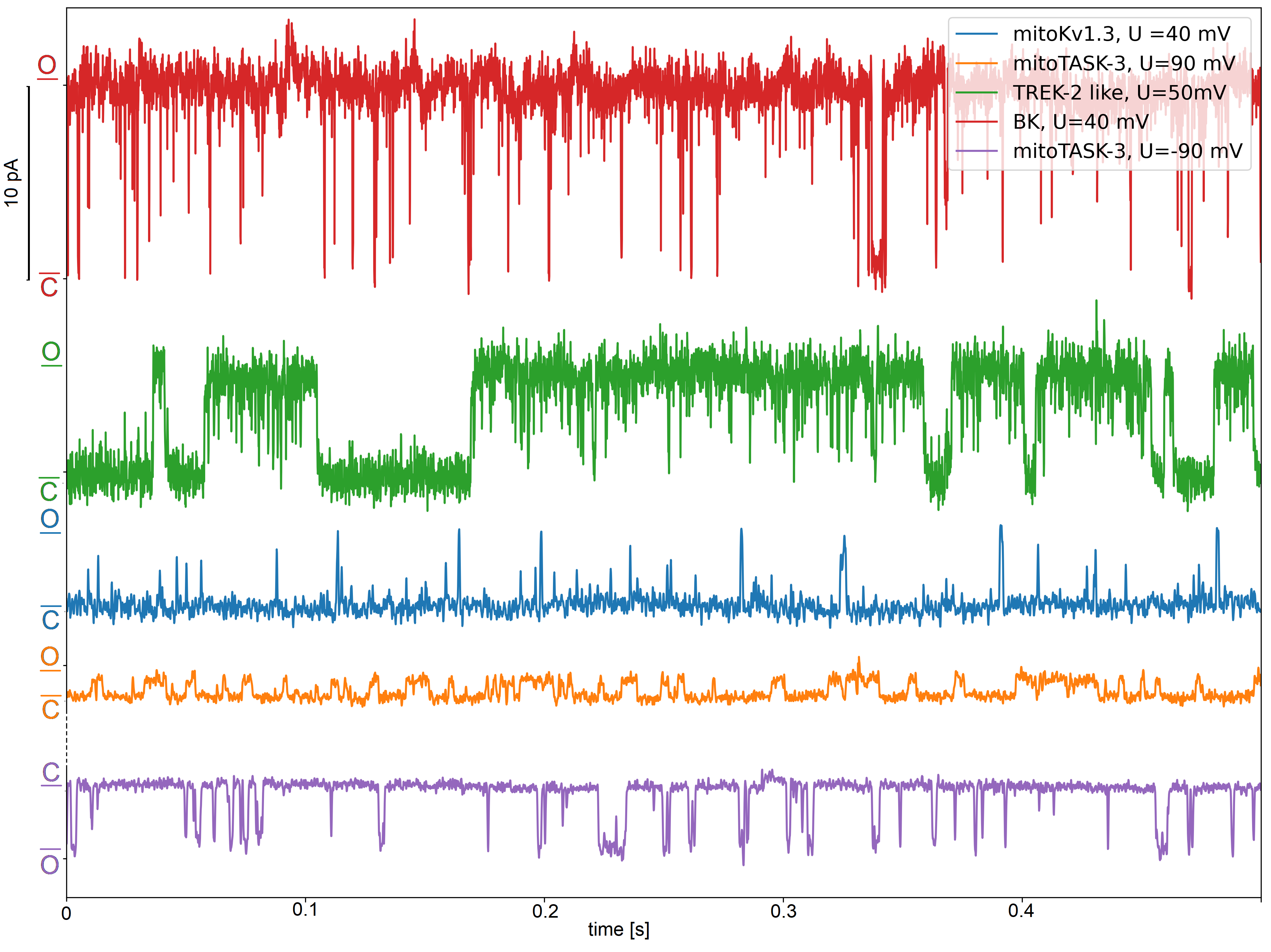}
    \caption{The samples of single-channel currents recorded for different types of potassium channels.\label{figA1}}
\end{figure}

\begin{figure}[H]
    \includegraphics[width=0.5\textwidth]{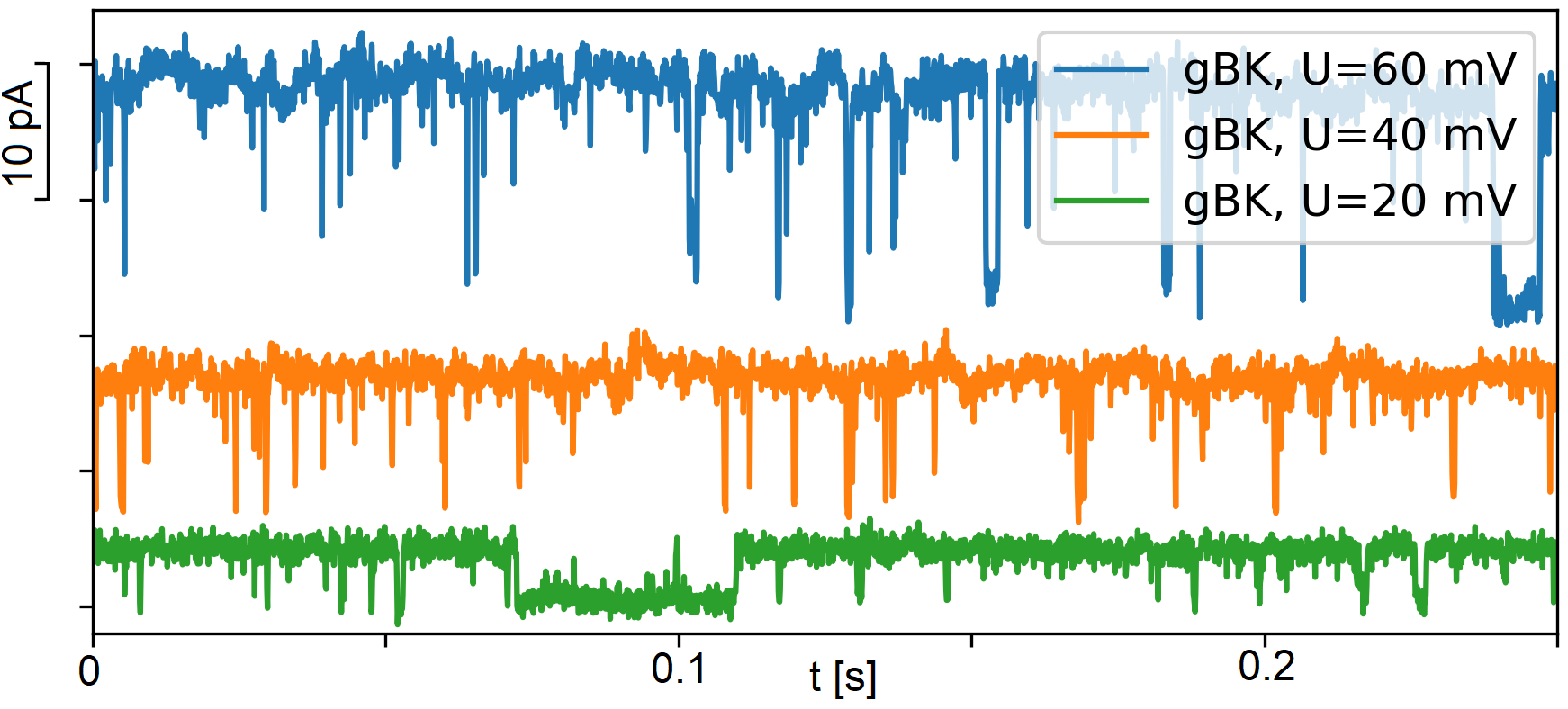}
    \includegraphics[width=0.5\textwidth]{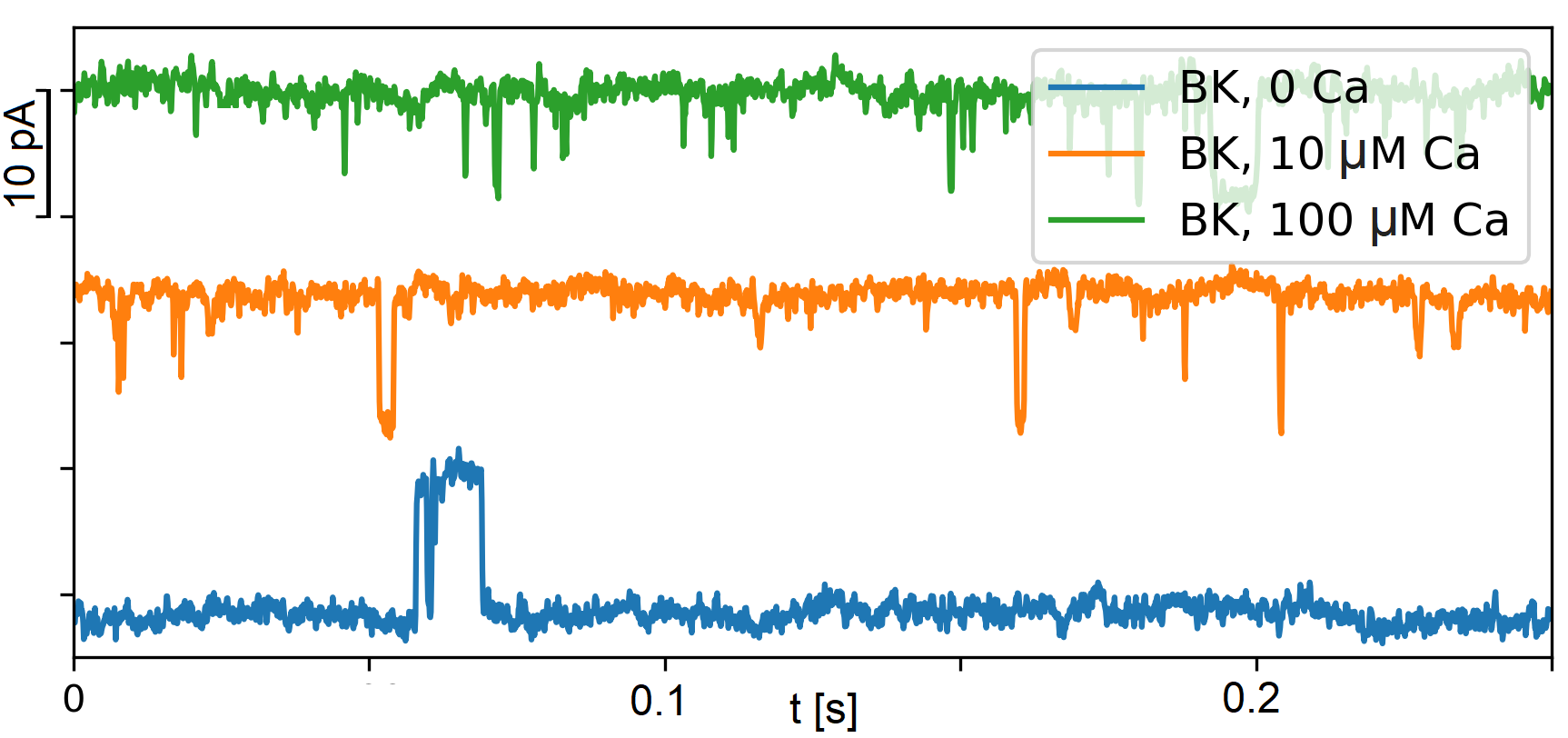}
    \caption{The samples of single-channel currents recorded for BK channels from glioblastoma cells (gBK) at different voltages (left panel) and from human bronchial epithelial cells at different concentrations of $Ca^{2+}$ ions.\label{figA2}}
\end{figure}

\begin{figure}[H]
    \includegraphics[width=0.5\textwidth]{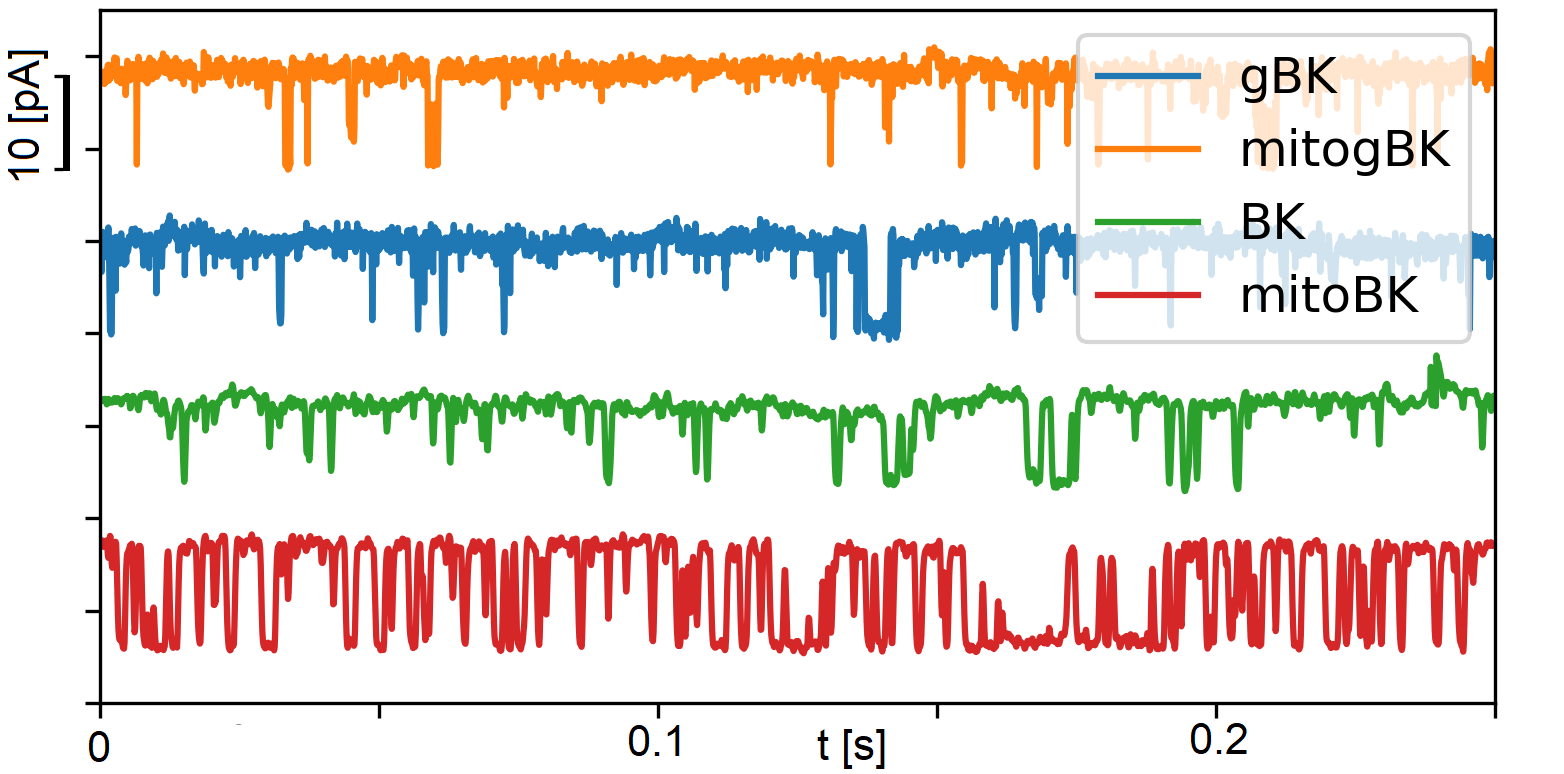}
    \caption{The samples of single-channel currents recorded for plasma membrane and mitochondrial BK channels from glioblastoma cells (gBK and mitoBK) and their analogs from endothelial cells (BK and mitoBK).\label{figA3}}
\end{figure}

\bibliographystyle{elsarticle-num} 
\bibliography{bibliography}





\end{document}